\documentstyle[epsfig,aps,prb,multicol]{revtex}
\begin{document}

\title{Charge redistribution and polarization energy of organic
molecular crystals}

\author{E.V. Tsiper and Z.G. Soos}

\address{Department of Chemistry, Princeton University, Princeton, NJ
08544}

\date{June 5, 2001}

\maketitle

\begin{multicols}{2}
 [ \begin{abstract}
 We present an approach to electronic polarization in molecular solids
treated as a set of quantum systems interacting classically.
Individual molecules are dealt with rigorously as quantum-mechanical
systems subject to classical external fields created by all other
molecules and, possibly, external sources.  Self-consistent equations
are derived for induced dipoles and for atomic charges whose
redistribution in external fields is given explicitly by an atom-atom
polarizability tensor.  Electronic polarization is studied in two
representative organic molecular crystals, anthracene and
perylenetetracarboxylic acid dianhydride (PTCDA), and contrasted to
previous results for systems of polarizable points.  The stabilization
energies of the neutral lattice, of isolated anions and cations, and
of cation-anion pairs are found.  Charge redistribution on ions is
included.  The dielectric tensors of crystals are successfully related
to gas-phase properties and provide consistency checks on polarization
energies.  The procedure is generally applicable to organic crystals
in the limit of no intermolecular overlap.
 \end{abstract} ]

\section{Introduction}

Recent advances in the preparation of ordered thin films\cite{forrest}
and of organic molecular crystals\cite{dodabalapur,batlogg} with reduced
impurity levels have revived interest in organic electronics and the
properties of charge carriers in these materials.  Organic molecular
crystals are typically insulators with low dielectric constant,
$\kappa\sim3$, and charges localized on molecules.  Electronic
polarization, an effect that is usually not considered in conventional
inorganic semiconductors, has a central role in the electronic
properties of organic crystals, as discussed by Gutmann and
Lyons,\cite{gutmann} Pope and Swenberg,\cite{pope} and Silinsh and
Capek.\cite{silinsh} When a charge carrier is brought into a molecular
solid, its field polarizes the surrounding molecules.  Secondary
polarization fields created by polarized molecules contribute to the
total self-consistent polarization cloud that surrounds each charged
quasiparticle.  Such a cloud is sometimes referred to as an electronic
polaron, to distinguish from lattice relaxation effects.

The overall energy relaxation of a positive and negative charge
carrier, $P_++P_-$, is an important property of an organic material.
Typical values\cite{gutmann,pope,silinsh} are in the range of 2---3
eV.  The transport gap, i.e. the energy necessary to create a
well-separated electron-hole pair is

\begin{equation}
E_t=I-A-P_+-P_-
\label{Et}
\end{equation}
 where $I$ and $A$ are the gas-phase ionization potential and electron
affinity of the molecule.  $P_+$ and $P_-$ contain polaronic
contributions due to both intramolecular and lattice phonons that are
estimated\cite{silinsh} to be $\sim10\%$.  We focus on the large {\em
electronic} components of $P_+$ and $P_-$ and from now on exclude
polaronic effects.

The appearance in Eq.~(\ref{Et}) of gas-phase properties is made
possible by weak intermolecular forces and Van der Waals
separations in organic molecular crystals.  The solid-state
environment is taken as a perturbation in molecular exciton theory.
In contrast to inorganic semiconductors, organic crystals can normally
be approximated as molecules with negligible overlap, and vanishing
intermolecular overlap is the crucial approximation to our general
development of electronic polarization energies.  The energy necessary
to create an electron-hole pair at finite separation ${\bf r}$ defines
the interaction potential $V({\bf r})<0$:

\begin{equation}
E_{\rm pair}({\bf r})=E_t+V({\bf r}),
\label{Epair}
\end{equation}
 Nearest-neighbor ion pairs of large molecules obviously deviate from
point charges.

Mott and Littleton\cite{mott} first estimated the electronic
polarization energy in an atomic lattice by considering each atom as a
polarizable point.  The self-consistent treatment of point charges in
lattices of polarizable points was subsequently developed by
Munn,\cite{munn} Silinsh,\cite{silinsh} and coworkers.  While
satisfactory for {\em atomic} lattices, polarizable points for
molecules completely neglect their structure.  Polarization
studies,\cite{silinsh,munn} primarily on the acene family, addressed
molecular size to some extent by introducing submolecules, whose
choice is arbitrary, that carry a fraction of the total molecular
polarizability.  A recent application\cite{reis} to anthracene has
several choices, the most elaborate one being submolecules at all
carbons, while a study\cite{petelenz} of perylenetetracarboxylic acid
dianhydride (PTCDA) has 11 submolecules centered at rings and CO
bonds.  In effect, the quantum nature of molecules is approximated by
the microelectrostatics of submolecules.

In large $\pi$-conjugated molecules subject to an external field,
charge redistributes over distances comparable to the size of the
molecule and generates large nonlinear optical responses.  Such a flow
of charge in molecules creates secondary polarization fields that do
not necessarily reduce to the field of a set of induced dipoles.
Thus, rigorous treatment of the polarization field requires analysis
of the molecular charge distribution $\rho({\bf r})$.

Another issue is electrostatic interactions present already in the
ground state of a condensed phase composed of neutral molecules,
especially when they contain heteroatoms.  These charges induce mutual
polarization in the surrounding molecules and contribute to the
overall stabilization energy of the solid.  We estimate that the
polarization contribution to the ground-state energy can be quite
significant, reaching hundreds of meV per molecule.  A notable
exception is the acene family, whose $\pi$-systems have approximate
electron-hole symmetry and negligible partial charges, making the
polarization energy small.

In this paper we present an approach to electronic polarization in
molecular solids that allows for quantitative description of
intramolecular charge redistribution.  The crucial approximation is
the neglect of intermolecular overlap.  Zero overlap implies that
coordinate space can be subdivided into non-overlapping regions, e.g.
Wigner-Seitz cells, associated with individual molecules.  Each
molecule is then a quantum-mechanical system subject to external
fields created by the crystal and, possibly, other external sources.
The external fields are rigorously {\em classical}, so that quantum
mechanics is needed at the intramolecular level only.

Section II illustrates the idea and presents self-consistent
equations.  Section III describes an approximate discrete form of the
equations that, we believe, strikes an optimum balance between
accuracy and simplicity for practical use.  In Sec.~IV the equations
are applied to the translationally-invariant lattice to find the
polarization contribution to the binding energy.  In Sec.~V equations
for the polarization energy of a system of ions embedded in a lattice
are derived.  The polarization energy of isolated charge carriers in
anthracene and PTCDA crystals is calculated, as well as the energy of
various ion pairs.  The dielectric tensors of anthracene and PTCDA are
computed in Sec.~VII, and their consistency with polarization energy
calculations is established.  We compare $P_+$, $P_-$, and $V({\bf
r})$ to submolecular results and comment briefly in Sec.~VIII on
applications to organic solids with mobile charge carriers.

\section{Self-consistent charge distribution}

In the zero-overlap approximation, the self-consistent solution of the
Schr\"odinger equation for a solid reduces to a product of wave
functions of individual molecules.  The minimum energy relative to
gas-phase (noninteracting) molecules or ions can be done in two steps.
First, the change $E(\phi)$ in the ground state energy of each
molecule is found as a functional of the external electrostatic
potential $\phi({\bf r})$.  For simplicity we neglect any magnetic
interactions.  The ground state charge distribution $\rho({\bf
r};\phi)$ also depends on $\phi$ and determines the secondary
polarization field created by the molecule.

The total electrostatic potential at a point ${\bf r}$ within a
molecule $a$ is created by all other molecules $b\neq a$ and,
possibly, an applied field:

\begin{equation}
\phi^a({\bf r})=\phi^a_{\rm appl}({\bf r})
 +{\sum_b}^{'}\!\int d^3r^{'}
  \frac{\rho^b({\bf r}^{'};\phi^b)}
       {|{\bf r}-{\bf r}^{'}|}
 \label{phitot}
\end{equation}
 The prime at the sum excludes the term with $b=a$.  The total energy
of the solid is then

\begin{equation}
E_{\rm tot}=\sum_aE^a(\phi^a)
  -{\sum_{a<b}}\!\int\!\int d^3rd^3r^{'}
  \frac{\rho^a({\bf r};\phi^a)\rho^b({\bf r}^{'};\phi^b)}
       {|{\bf r}-{\bf r}^{'}|}
 \label{EtotG}
\end{equation}
 The second term compensates for double-counting the intermolecular
interactions in the first term.  Equivalently,

\begin{equation}
E_{\rm tot}=\sum_a\left[E^a
  -\frac{1}{2}\int d^3r
  (\phi^a-\phi^a_{\rm appl})\rho^a
  \right]
 \label{EtotG1}
\end{equation}
 Minimization of $E_{\rm tot}$ with respect to $\phi^a({\bf r})$ yields
the self-consistent ground state energy of the molecular solid in the
approximation of zero overlap.

In principle, variation of Eq.~(\ref{EtotG1}) with respect to $\phi^a$
gives an equation for $\rho^a({\bf r})$, which together with
Eq.~(\ref{phitot}) forms a complete self-consistent system.  A more
efficient way, perhaps, is to use an appropriate quantum chemical
procedure to find $\rho^a({\bf r};\phi^a)$ for each molecule $a$ and
iterate several times by updating $\phi^a({\bf r})$ using
Eq.~(\ref{phitot}).  The self-consistent problem defined by
Eqs.~(\ref{phitot}) and (\ref{EtotG1}) for charges and potentials is
typical of classical electrostatics.  The quantum part is limited to
the charge distribution $\rho^a({\bf r},\phi^a)$.

The index $a$ can be dropped or restricted to a single unit cell when
the translationally-invariant state of the lattice is of interest.
This is the case, for example, in the calculation of the dielectric
tensor.\cite{ours} Otherwise, a finite number of molecules must be
considered.  A practical implementation of the procedure requires some
form of discretization of the continuous functions $\phi^a({\bf r})$
and $\rho^a({\bf r})$ defined within the molecular volume.  Certain
trade-off between the accuracy and simplicity suitable for repetitive
quantum-chemical calculation is unavoidable.  In the following section
we develop a simple scheme, which captures intramolecular charge
redistribution to a quantitative accuracy.  The procedure has been
successfully implemented by us to calculate indices of refraction of
anthracene and PTCDA.\cite{ours}

\section{Molecule in nonuniform field}

In this section we omit the index $a$ and consider a single molecule
subject to an external potential $\phi({\bf r})$.  We note that charge
redistribution gives a major contribution to the polarizability of
large conjugated molecules.  This ``major part'' is not defined
quantitatively, as there is no unique definition of atomic charges.
The scheme we develop below separates molecular polarizability into
two parts, the sum of which matches the actual molecular
polarizability $\alpha$ with the best value known from experiment or
theory.

We use a semiempirical Hamiltonian because it provides a natural way
to represent an arbitrary external potential acting on a molecule.  We
define $\phi_i=\phi({\bf r}_i)$, the potential at the position of each
atom ${\bf r}_i$.  A site energy $\phi_i$ is added to the diagonal
matrix elements for the orthogonalized valence orbitals of atom $i$.
We employ the INDO/S Hamiltonian,\cite{indos} which is known to
approximate molecular properties at only a tiny fraction of the cost
of {\em ab-initio} calculations.  Throughout the paper we use L\"owdin
charges $\rho_i^a$, where $i$ labels atoms in molecule $a$.  The
charges are defined as the sums of occupation numbers of
orthogonalized orbitals of atom $i$.

The corresponding contribution $\alpha^C$ to the actual polarizability
$\alpha$ is clearly restricted to the molecular plane in conjugated
molecules.  We associate the difference $\alpha-\alpha^C$ between the
actual and INDO/S polarizabilities with ``atomic'' contributions
caused by the distortion of atomic orbitals in the field.  Atomic
contributions are small corrections to large in-plane
polarizabilities.  We note that {\em any} choice of $\rho_i(\bbox{F})$
leads to in-plane $\alpha^C$ as a consequence of a discrete charge
distribution.

Atomic contributions to $\alpha$ can be described, as in atomic
lattices, in terms of induced dipoles situated at the positions of
atoms.  Based on such an idea we propose the following minimal scheme
that is designed to capture both charge-redistribution and ``atomic''
parts of the molecular response to external fields.  We describe the
state of an $N$-atom molecule by a set of $2N$ variables, $\rho_i$ and
${\bbox{\mu}_i}$, which represent the partial charges and induced
dipole moments of atoms.  The same number of variables,
$\phi_i=\phi({\bf r}_i)$ and $\bbox{F}_i=-\bbox{\nabla}\phi({\bf
r}_i)$, describes the external field acting on a molecule.

We denote by $q_i$ the deviation of partial charges from the
ground-state values $\rho_i^{(0)}$ of an isolated molecule or
molecular ion,

\begin{equation}
q_i(\phi)=\rho_i(\phi)-\rho_i^{(0)}
\end{equation}
 At small fields the energy of the molecule is quadratic in the
distortion from equilibrium:

\begin{eqnarray}
E(\rho_i,\bbox{\mu}_i;\phi_i,\bbox{F}_i)&=&\frac{1}{2}\sum_{ij}q_i\Pi^{-1}_{ij}q_j+\sum_i\rho_i\phi_i\nonumber\\
 &+&\frac{1}{2}\sum_i\bbox{\mu}_i\widetilde\alpha^{-1}_i\bbox{\mu}_i
 -\sum_i\bbox{\mu}_i\bbox{F}_i
 \label{EmolF}
\end{eqnarray}
 Here the positive-definite charge stiffness matrix $\Pi^{-1}$
describes the increase in the internal energy of the molecule when the
charge distribution deviates form its zero-field equilibrium; the
tensor $\widetilde\alpha_i^{-1}$ plays the same role for atomic
dipoles.  At given configuration $\{\phi_i,\bbox{F}_i\}$ of the
external field, the minimum of the energy functional Eq.~(\ref{EmolF})
is achieved at

\begin{mathletters}
\begin{eqnarray}
\rho_i&=&\rho_i^{(0)}-\sum_j\Pi_{ij}\phi_j
\label{SC1a}\\
\bbox{\mu}_i&=&\widetilde\alpha_i\bbox{F}_i,
\label{SC1b}
\end{eqnarray}
\label{SC1}
\end{mathletters}
 We see that $\widetilde\alpha_i$ is the polarizability for atom $i$.
It does not necessarily reduce to a scalar, since atoms in a molecule
have no rotational symmetry.  We could also assume nonzero atomic
dipoles $\bbox{\mu}_i^{(0)}$ in the ground state of nonpolar molecules
and so obtain symmetric equations, but we set $\bbox{\mu}_i^{(0)}=0$
in this paper.

The energy of the molecule at the minimum is

\begin{equation}
E(\phi,\bbox{F})=\sum_i\rho_i^{(0)}\phi_i
  +\frac{1}{2}\sum_i\left(q_i\phi_i-\bbox{\mu}_i\bbox{F}_i\right).
\label{Emol}
\end{equation}
 Equivalently,

\begin{equation}
E(\phi,\bbox{F})=\sum_i\rho_i^{(0)}\phi_i
  -\frac{1}{2}\sum_{ij}\phi_i\Pi_{ij}\phi_j
  -\frac{1}{2}\sum_i\bbox{F}_i\widetilde\alpha\bbox{F}_j.
\label{Emol1}
\end{equation}
 The last two terms describe the energy relaxation of the molecule in
the external field.  The positive-definite symmetric matrix $\Pi_{ij}$
is the susceptibility with respect to site potentials $\phi_i$
[cf. Eqs.~(\ref{SC1a}) and (\ref{Emol1})]:

\begin{equation}
\Pi_{ij}=-\left(\frac{\partial\rho_i}{\partial\phi_j}\right)_0
   =-\left(\frac{\partial^2E}{\partial\phi_i\partial\phi_j}\right)_0
\label{Pi}
\end{equation}
 Partial derivatives are evaluated at $\phi_i=0$.  $\Pi_{ij}$
determines the charge redistribution among atoms in the external
potential.  It is a natural extension of the similar quantity
$\pi_{ij}$ used in $\pi$-electron theory,\cite{coulson} and called the
atom-atom polarizability.  In our case, the total charge of all
valence electrons is considered.  Note that our definition differs by
a factor $-1/2$.

Atom-atom polarizabilities $\Pi_{ij}$ obey the condition

\begin{equation}
\sum_i\Pi_{ij}=\sum_j\Pi_{ij}=0,
\label{Picond}
\end{equation}
 since the charge distribution in Eq.~(\ref{SC1a}) is invariant to an
additive constant in all site potentials $\phi_i$.  The zero-overlap
approximation conserves charge at each molecule.  Expansion to second
order in $\phi_i$ is sufficient for $\phi_i<1$ eV.  Eqs.~(\ref{SC1})
eliminate the need to solve repetitively the quantum problem for the
molecule.  We calculate $N(N+1)/2$ atom-atom polarizabilities
$\Pi_{ij}$ only once using Eq.~(\ref{Pi}) for the neutral molecule and
for the cation and anion.  Stronger perturbations may require
re-evaluation of $\Pi_{ij}$ at some intermediate $\phi_i$.

The total induced moment of a molecule is

\begin{equation}
\bbox{\mu}=\sum_i\left({\bf r}_i\rho_i+\bbox{\mu}_i\right).
\label{mu}
\end{equation}
 The molecular polarizability consists, therefore, of two terms,
$\alpha=\alpha^C+\widetilde\alpha$:

\begin{equation}
\alpha^{\alpha\beta}=\sum_{ij}\Pi_{ij}r_i^\alpha r_j^\beta
  +\sum_i\widetilde\alpha_i^{\alpha\beta}.
\label{alpha}
\end{equation}
 where the Greek indices take the values $x$, $y$, and $z$.

Equation (\ref{alpha}) illustrates the advantages and limitations of
partial atomic charges.  With the aid of $\Pi_{ij}$, they provide a
rigorous description of charge redistribution.  The assumption of
polarizable points is atomic lattices is kept, however, through
$\widetilde\alpha=\alpha-\alpha^C$.  We have corrections to INDO/S
charges and distribute $\widetilde\alpha$ proportionally to the
numbers of valence electrons $n_i$ associated with individual atoms:
$\widetilde\alpha_i=\widetilde\alpha n_i/\sum n_i$.  As in previous
theory,\cite{silinsh,mott,munn,reis,petelenz} $\alpha$ is an
independent gas-phase input to the calculation.

 \vskip 0.2 in
 {\small {\bf Table~I} Principal components of the molecular
polarizabilities of anthracene and PTCDA; the long (L), medium (M),
and normal (N) axes are fixed by D$_{2h}$ symmetry.}

\noindent
\begin{tabular}{lccc}
\\
\tableline
\tableline
Method &
$\alpha_{NN}$ (\AA$^3$) &
$\alpha_{MM}$ (\AA$^3$) &
$\alpha_{LL}$ (\AA$^3$)
\\
\tableline
\\
\multicolumn{2}{l}{\bf \ \ \ \ Anthracene} & & \\
Experiment\protect\cite{cheng_lefevre} &
                        15.2      & 25.6      & 35.2  \\
                      & 15.9      & 24.5      & 35.9  \\
B3LYP/6-311++G**      & 12.03     & 24.27     & 42.56 \\
INDO/S ($\alpha^C$)   & 0         & 24.05     & 41.52 \\
\\
\multicolumn{2}{l}{\bf \ \ \ \ PTCDA} & & \\
B3LYP/6-311++G**      & 18.06     & 50.27     & 88.18 \\
INDO/S ($\alpha^C$)   & 0         & 50.84     & 84.54 \\
\tableline
\tableline
\\
\end{tabular}

Table~I summarizes results for anthracene and PTCDA.  Density
functional (B3LYP) results have been obtained using the Gaussian 98
program.\cite{gaussian} Theory and experiment are in reasonable
agreement for anthracene molecules when large basis sets are
used,\cite{ours,reis} (such as 6-311++G**).  Dielectric
data\cite{forrest_apl} for crystalline PTCDA films are also
consistent\cite{ours} with calculated molecular polarizabilities.  The
INDO/S results for $\alpha^C$ from Eq.~(\ref{alpha}) are confined to
in-plane components that represent charge redistribution according to
$\Pi_{ij}$ in Eq.~(\ref{alpha}).  We also need
$\widetilde\alpha=\alpha-\alpha^C$.  Unless otherwise indicated, we
will use the B3LYP polarizabilities in Table~I.  Since they exceed
$\alpha^C$, atomic contributions increase the polarization compared to
the ``charges-only'' choice of $\widetilde\alpha=0$.  We note that
simple H\"uckel theory often overestimates responses to applied fields
and hence the amount of charge redistribution; in that case
$\widetilde\alpha$ may be negative and the atomic part reduces the
polarization.  Equations (\ref{SC1}) hold for any $\widetilde\alpha$.

\section{Self-consistent equations}

In the condensed phase the potential and field at the position of atom
$i$ of molecule $a$ created by all other molecules $b\neq a$ are

\begin{mathletters}
\begin{eqnarray}
\phi_i^a&=&{\sum_b}^{'}\sum_j
   v({\bf r}_{ij}^{ab})\rho_j^b
  +v_\beta({\bf r}_{ij}^{ab})\mu_j^{b\beta},\\
F_i^{a\alpha}&=&{\sum_b}^{'}\sum_j
   v_\alpha({\bf r}_{ij}^{ab})\rho_j^b
  +v_{\alpha\beta}({\bf r}_{ij}^{ab})\mu_j^{b\beta},
\end{eqnarray}
\label{SC2}
\end{mathletters}
 where $v({\bf r})=1/r$, $v_\alpha({\bf r})=-\partial v/\partial
r^\alpha$, $v_{\alpha\beta}({\bf r})=\partial^2 v/\partial
r^\alpha\partial r^\beta$.  Summation over repeated Greek indices
$\alpha,\beta=x,y,z$ is assumed.  The vector ${\bf r}_{ij}^{ab}={\bf
r}_i^a-{\bf r}_j^b$ points to the atom of interest from atom $j$ of
molecule $b$.  Here we have assumed no external sources for
simplicity.  Equations (\ref{SC2}) together with Eqs.~(\ref{SC1}) form
a complete self-consistent linear system for $\rho_i^a$,
$\bbox{\mu}_i^a$, $\phi_i^a$, and $\bbox{F}_i^a$.

The total polarization energy of the solid is [compare to
Eq.~(\ref{EtotG1})]

\begin{equation}
E_{\rm tot}=\sum_a
  \left[E^a
  -\frac{1}{2}\sum_i(\rho_i^a\phi_i^a-\bbox{\mu}_i^a\bbox{F}_i^a)
  \right].
\label{EtotF}
\end{equation}
 After some algebra and using Eq.~(\ref{Emol}), this reduces to

\begin{equation}
E_{\rm tot}=\frac{1}{2}\sum_a\sum_i\rho_i^{a(0)}\phi_i^a.
\label{Etot}
\end{equation}
 The derivation of this formula is simplified by replacing the dipoles
by pairs of charges separated by small distances, and taking the limit
in the final expression.

The total energy is a bilinear form of unperturbed charges
$\rho_i^{a(0)}$ and self-consistent potentials $\phi_i^a$.  We can
write it also in terms of self-consistent charges $\rho_i^a$ and
dipoles $\bbox{\mu}_i^a$, defining the unperturbed potentials
$\phi_i^{a(0)}$ and fields $\bbox{F}_i^{a(0)}$ in Eqs.~(\ref{SC2}) by
setting $\rho_i^a=\rho_i^{(0)}$ and $\bbox{\mu}_i^a=0$.  Using the
identity

\begin{equation}
\sum_{ai}\rho_i^{a(0)}\phi_i^a
  =\sum_{a\neq b}\sum_{ij}v({\bf r}_{ij}^{ab})\rho_i^{a(0)}\rho_j^b
  =\sum_{bj}\rho_j^{b}\phi_j^{b(0)},
\end{equation}
 the total energy becomes

\begin{equation}
E_{\rm tot}=\frac{1}{2}\sum_a\sum_i(\rho_i^a\phi_i^{a(0)}-
   \bbox{\mu}_i^a\bbox{F}_i^{a(0)}).
\label{Etot1}
\end{equation}
 We use this form of $E_{\rm tot}$ in Sec.~VI to treat ions in
infinite lattices.

Equation (\ref{Etot1}) reduces to the previous
result\cite{fox,silinsh,munn} when all molecules are shrunk to points,
with $\sum_i\rho_i^a=0$ for molecules and $\pm1$ for ions.  Since
charge redistribution is no longer possible, we may set $\Pi_{ij}=0$,
$\alpha^C=0$, and polarizability $\widetilde\alpha=\alpha$ at
positions of neutral molecules.  The first correction to $\alpha^C$
for finite molecules is an induced dipole at the center, which gives
the same approximation for $\alpha$ when combined with
$\widetilde\alpha$.  The potential $\phi_i^{a(0)}$ in
Eq.~(\ref{Etot1}) is due to ions, and the first sum, which is now
restricted to charged sites, becomes the source term $W_0$ of
Ref.~\onlinecite{fox}.  The second term, over molecules, describes
induced dipoles in the field of the ions and is the $W_1$ term of
Ref.~\onlinecite{fox}.  The polarizability of ions is generally
different from molecules, but is not required for finding $P_\pm$ in
centrosymmetric lattices of point molecules, since the ion is at an
inversion center.

The expression (\ref{Etot}) or (\ref{Etot1}) for the lattice
polarization energy is not restricted to equivalent molecules.  In
principle, each molecule $a$ may have its own $\rho_i^{a(0)}$,
$\Pi_{ij}^a$, and $\widetilde\alpha_i^a$.  In practice, there are
several molecules per unit cell in organic molecular crystals.  The
translationally-invariant lattice of neutral molecules, the neutral
lattice of the following section, reduces to atomic charges and
potentials within a unit cell.  Molecular ions in specified unit cells
break translational symmetry and, as discussed in Secs.~VI and VII,
require different methods for finding $E_{\rm tot}$.  In the
zero-overlap approximation, charge carriers are molecular ions in
place of neutral molecules.  The polarization energy $P_\pm$ of a
carrier is the energy difference between two extensive quantities, the
lattice with the ion and the neutral lattice.

\section{Neutral lattice}

In this section we evaluate the polarization energy of the neutral
lattice.  The analogous quantity vanishes identically in the
polarizable-point approach, since there are no fields or induced
dipoles in the lattice until charges are introduced.  The
self-consistent Eqs.~(\ref{SC1}), (\ref{SC2}) can be restricted to a
single unit cell of volume $v_c$.  The problem, therefore, reduces to
a system of $4NN_c$ linear equations, where $N_c$ is the number of
molecules in a unit cell, and $N$ is the number of atoms in a
molecule.

Madelung-type infinite sums in Eqs.~(\ref{SC2}) can be evaluated using
Ewald's method.\cite{madelung} Special care has to be taken to treat
complex lattices with many partial charges in a unit cell.  For this
purpose we introduce a fictitious uniform neutralizing background for
each partial charge.  Since the unit cell is neutral, these
backgrounds cancel exactly.  We define an auxiliary potential function
${\cal V}({\bf r})$,

\begin{equation}
{\cal V}({\bf r})=\lim_{R\rightarrow\infty}
  \left[{\sum_{|\bbox{\ell}|<R}}^{'}\frac{1}{|{\bf r}-\bbox{\ell}|}
  -\frac{2\pi r^2}{3v_c}
  -\left(\frac{9\pi M_R^2}{2v_c}\right)^{1/3}
  \right]
\end{equation}
 where the summation is over $M_R$ lattice vectors $\bbox{\ell}$
falling within a sphere of radius $R$, with the term $\bbox{\ell}=0$
missing.  Subtracted is the potential of the uniform neutralizing
spherical charge $-M_R$ centered at the origin of coordinates.
Centering of the neutralizing backgrounds at a common point in space
is necessary for proper cancelation.

Ewald's method gives

\end{multicols}

\begin{equation}
{\cal V}({\bf r})=
   -\frac{2\pi r^2}{3v_c}
   -\frac{\text{erf}(Gr)}{r}+
   {\sum_{\bbox{\ell}}}^{'}
      \frac{1-\text{erf}(G|{\bf r}-\bbox{\ell}|)}{|{\bf r}-\bbox{\ell}|}
   -\frac{\pi}{v_cG^2}
   +\frac{\pi}{v_cG^2}{\sum_{{\bf g}}}^{'}
      \frac{\exp(-g^2/4G^2)}{g^2/4G^2}\cos({\bf g}{\bf r}),
\label{Vcal}
\end{equation}

\begin{multicols}{2}

 \noindent\ 
 where $\text{erf}(x)=(2/\sqrt{\pi})\int_0^x dy\exp(-y^2)$ is the
error function.  The second sum is over all reciprocal vectors ${\bf
g}\neq0$, $\exp(i{\bf g}\bbox{\ell})=1$.  Ewald's parameter $G$ is
arbitrary (the result does not depend on its value); a reasonable
choice is $G=(\pi^2/v_c)^{1/3}$.  The function ${\cal V}({\bf r})$ is
regular within the central lattice cell $\bbox{\ell}=0$, including the
point ${\bf r}=0$, and it is not periodic because of the missing term.
The function ${\cal V}({\bf r})+1/r$ is periodic in ${\bf r}$.

Equations (\ref{SC2}) can be written in terms of ${\cal V}({\bf r})$
and its derivatives ${\cal V}_\alpha({\bf r})=-\partial{\cal V}({\bf
r})/\partial r^\alpha$ and ${\cal V}_{\alpha\beta}({\bf
r})=\partial^2{\cal V}({\bf r})/\partial r^\alpha\partial r^\beta$:

\begin{mathletters}
\begin{eqnarray}
\phi_i^a&=&
{\sum_b}^{'}\sum_j
    v({\bf r}_{ij}^{ab})\rho_j^b
   +v_\beta({\bf r}_{ij}^{ab})\mu_j^{b\beta}\nonumber\\
  &+&\sum_b\sum_j
    {\cal V}({\bf r}_{ij}^{ab})\rho_j^b
   +{\cal V}_\beta({\bf r}_{ij}^{ab})\mu_j^{b\beta},
\label{SC2ta}\\
F_i^{a\alpha}&=&
{\sum_b}^{'}\sum_j
    v_\alpha({\bf r}_{ij}^{ab})\rho_j^b
   +v_{\alpha\beta}({\bf r}_{ij}^{ab})\mu_j^{b\beta}\nonumber\\
  &+&\sum_b\sum_j
    {\cal V}_\alpha({\bf r}_{ij}^{ab})\rho_j^b
   +{\cal V}_{\alpha\beta}({\bf r}_{ij}^{ab})\mu_j^{b\beta}.
\label{SC2tb}
\end{eqnarray}
\label{SC2t}
\end{mathletters}
 The sums over $b$ are restricted to the central unit cell.  In the
primed sums the term $b=a$ is excluded.  The terms with ${\cal V}$ and
its derivatives give contributions by charges and dipoles beyond the
central cell.

Equations (\ref{SC2t}) express $\phi_i^a$ and $\bbox{F}_i^a$ in terms
of $\rho_j^b$ and $\bbox{\mu}_j^b$.  Together with Eqs.~(\ref{SC1})
they form a complete linear system of $4NN_c$ equations, half of them
vector.  For example, $NN_c=48$ and 76, respectively, for anthracene
and PTCDA lattices, which results in 192 and 304 scalar linear
equations for these materials.

The solution for the neutral lattice is further denoted as
$\overline\rho_i^a$, $\overline{\bbox{\mu}}_i^a$, $\overline\phi_i^a$,
and $\overline{\bbox{F}}_i^a$.  These quantities are summed over
molecules in Eq.~(\ref{Etot}) or (\ref{Etot1}) to yield the
(extensive) self-consistent energy of the neutral lattice.

\subsection{Anthracene and PTCDA}

Using the procedure described above we calculated polarization energy
of anthracene and PTCDA crystals, which represent two major families
of organic semiconductors.  Both materials are monoclinic with two
molecules per unit cell.  Both molecules have centers of inversion and
are nearly planar in the crystal.  PTCDA molecules are co-planar, up
to a small tilt, and form layers and stacks.  In anthracene the angle
between molecular planes is significant.

We used the X-ray crystal structures for PTCDA\cite{karl} and for
anthracene\cite{dunitz}.  The positions of hydrogens, not given
accurately by X-ray, were AM1-optimized using Gaussian.\cite{gaussian}
The gas-phase polarizability, needed to determine the atomic
correction $\widetilde\alpha$, is given in Table~I.

We obtain polarization energy of 330 meV per PTCDA molecule. This is
two orders of magnitude greater than 2.8 meV that we get for
anthracene.  The large polarization energy of the PTCDA lattice is
caused by significant partial atomic charges in neutral PTCDA
molecules, which are negligible in anthracene due to the approximate
electron-hole symmetry, as mentioned above.

\centerline{\epsfig{file=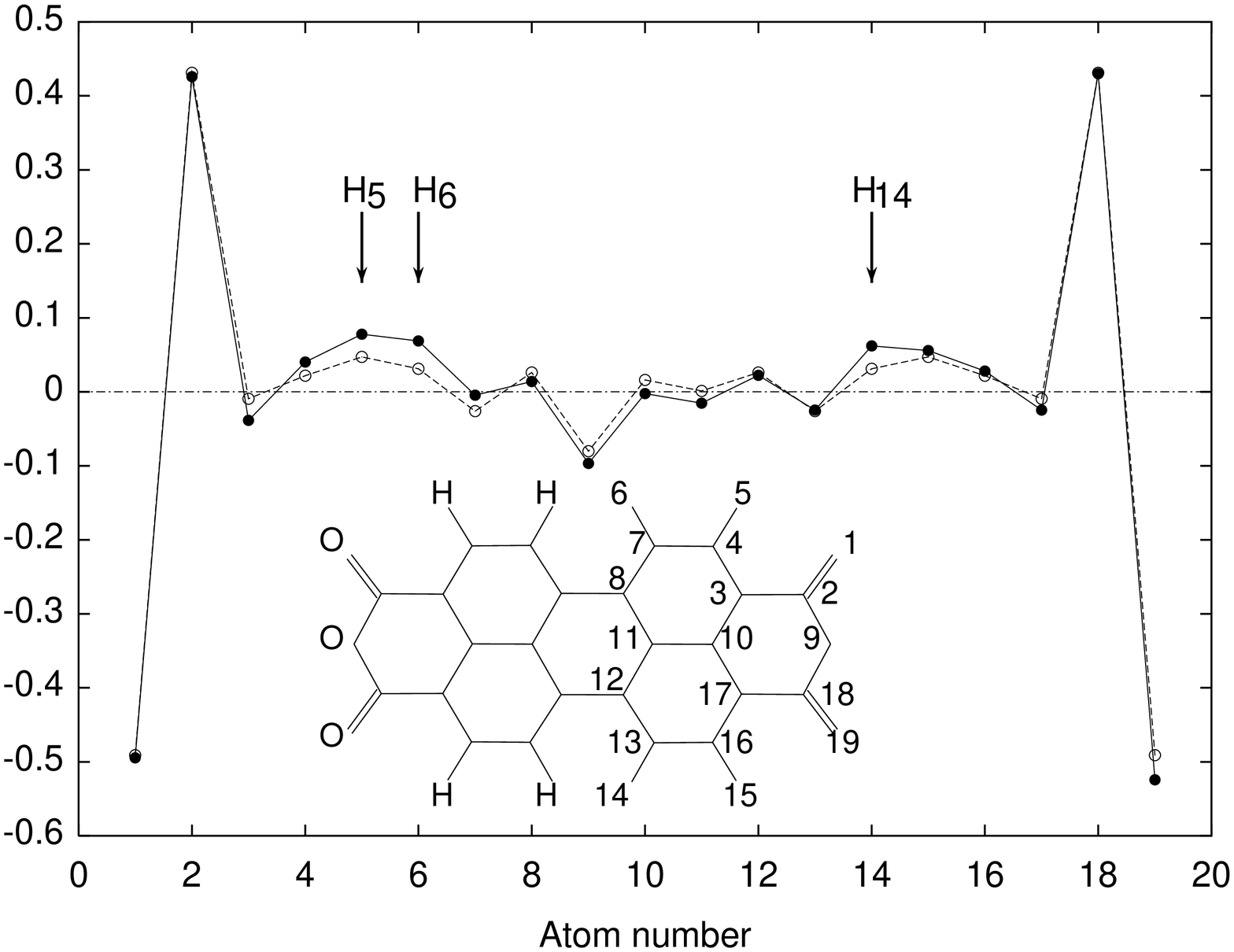,width=6cm,angle=-90}}
 {\small {\bf Fig.~1} Partial charges in PTCDA molecule in crystal
lattice ($\rho_i$, solid line) and in gas phase ($\rho_i^{(0)}$,
dashed line)}
 \vskip 0.1 in

Figure 1 compares partial atomic charges of a PTCDA molecule in the
gas phase and crystal.  The inset explains the atom numbering scheme.
Only one half of the molecule is shown because of C$_i$ symmetry.
Charge redistribution yields excess positive charge on three hydrogens
whose partial charges roughly double.  It is worth noticing that these
hydrogens reside in positions that suggest the formation of incipient
hydrogen bonds; the approximation of zero overlap excludes any
covalent contribution.  The distances from CH carbons to the nearest
oxygen atoms in neighboring molecules are 3.338, 3.269, and 3.768 \AA\
for C$_5$---O$_{1'}$, C$_{14}$---O$_{19'}$, and C$_6$---O$_{19'}$,
respectively, while the corresponding C---H---O angles are 
143.8, 149.3, and 158.4$^o$ (see Fig.~2).

\centerline{\epsfig{file=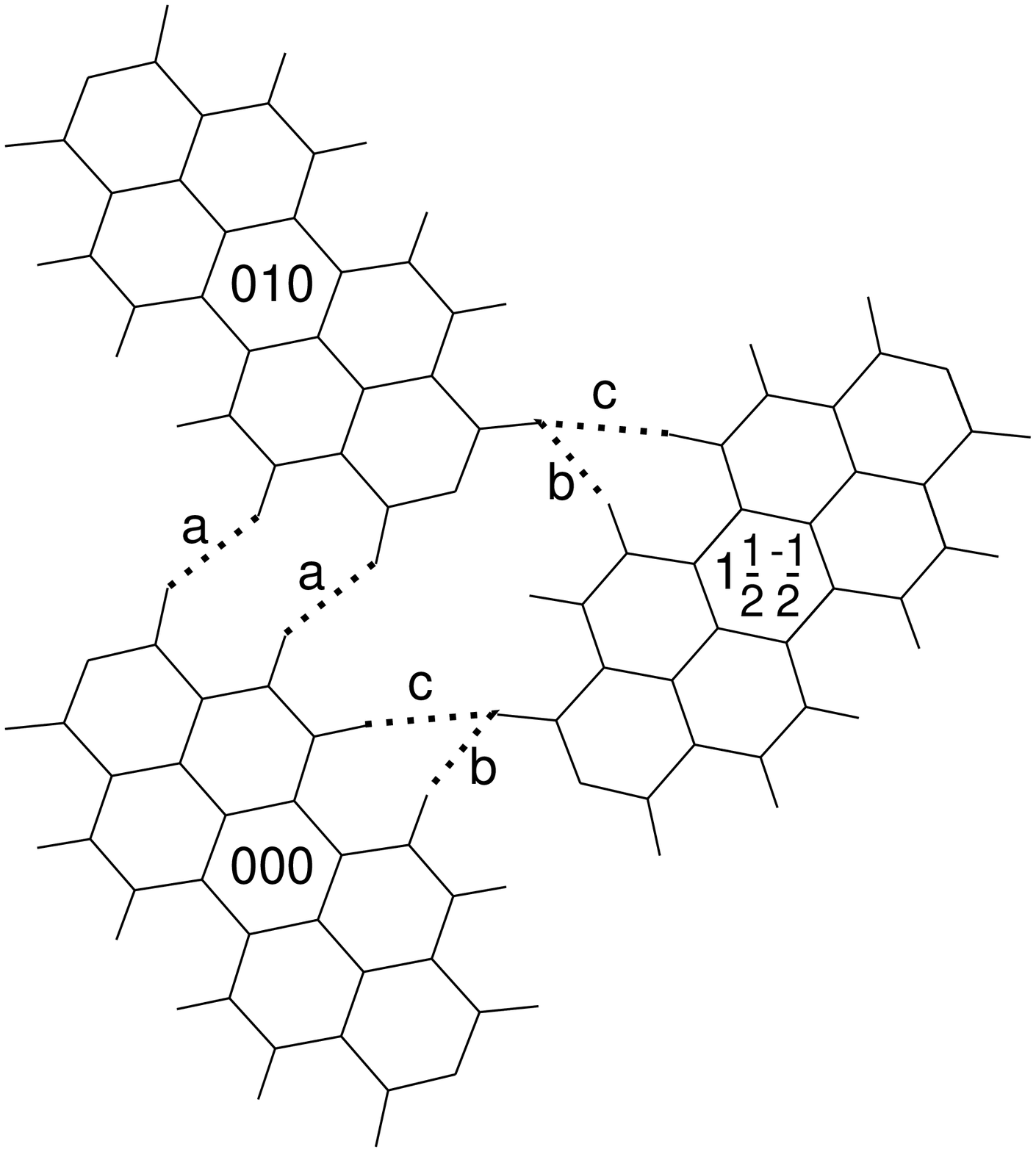,width=6cm}}
 {\small {\bf Fig.~2} Arrangement of PTCDA molecules in a layer
[projection onto (102) crystalline plane].  Incipient hydrogen bonds
(dotted lines) involve hydrogen atoms 5(a), 14(b), and 6(c), according
to the numbering.  Crystalline coordinates here and in Table~II
conform to the notation of Ref.~\onlinecite{karl}.}
 \vskip 0.1 in

\section{Polarization energy of charge carriers}

We now consider a lattice with one or more neutral molecules replaced
with molecular ions.  To evaluate the energy, we solve self-consistent
equations with the ions and subtract the polarization energy of the
neutral lattice. Each ion is described by $\Pi_{ij}^{ion}$,
$\widetilde\alpha^{ion}$, and $\rho_i^{ion(0)}$
($\sum\rho_i^{ion(0)}=\pm1$), which may differ from the similar
quantities in the neutral molecule.  While $\Pi_{ij}^{ion}$ and
$\rho_i^{ion(0)}$ are determined by semiempirical calculation, the
atomic correction $\widetilde\alpha^{ion}$ depends on a separate
calculation for the ion.  In this work, we use the same
$\widetilde\alpha$ for molecules and ions.

It is useful to rewrite the self-consistent equations (\ref{SC1}) and
(\ref{SC2}) in terms of the deviation from the neutral-lattice
solution:

\begin{mathletters}
\begin{eqnarray}
\delta\rho_i^a&=&\rho_i^a-\overline\rho_i^a,
\ \ \ \ \ \ \ 
\delta\phi_i^a=\phi_i^a-\overline\phi_i^a,\\
\delta\bbox{\mu}_i^a&=&\bbox{\mu}_i^a-\overline{\bbox{\mu}}_i^a,
\ \ \ \ \ 
\delta\bbox{F}_i^a=\bbox{F}_i^a-\overline{\bbox{F}}_i^a,
\end{eqnarray}
\end{mathletters}
 The equations for $\delta\rho_i^a$, $\delta\phi_i^a$, and each
component of $\delta\bbox{\mu}_i^a$ and $\delta\bbox{F}_i^a$ then read

\begin{mathletters}
\begin{eqnarray}
\delta\rho_i^a&=&\rho_i^{a*}-\sum_j\Pi_{ij}^a\delta\phi_j^a,
\label{SC3a}\\
\delta\mu_i^{a\alpha}&=&\widetilde\alpha_i^{\alpha\beta}\delta F_i^{a\beta},
\label{SC3b}
\end{eqnarray}
\label{SC3}
\end{mathletters}
 and

\begin{mathletters}
\begin{eqnarray}
\delta\phi_i^a&=&{\sum_b}^{'}\sum_j
   v({\bf r}_{ij}^{ab})\delta\rho_j^b
  +v_\beta({\bf r}_{ij}^{ab})\delta\mu_j^{b\beta},\\
\label{SC4a}
\delta F_i^{a\alpha}&=&{\sum_b}^{'}\sum_j
   v_\alpha({\bf r}_{ij}^{ab})\delta\rho_j^b
  +v_{\alpha\beta}({\bf r}_{ij}^{ab})\delta\mu_j^{b\beta}.
\label{SC4b}
\end{eqnarray}
\label{SC4}
\end{mathletters}
 The source term $\rho^{a*}$ is zero except for ions,

\begin{equation}
\rho_i^{ion*}=
  \Delta\rho_i^{ion(0)}-\sum_j\Delta\Pi_{ij}^{ion}\overline{\phi}_j^{ion}
  \equiv\delta\rho_i^{ion}(\overline{\phi}),
\label{rhostar}
\end{equation}
 where $\Delta\rho_i^{ion(0)}=\rho_i^{ion(0)}-\rho_i^{(0)}$,\ \
$\Delta\Pi_{ij}^{ion}=\Pi_{ij}^{ion}-\Pi_{ij}$, and
$\overline{\phi}_j^{ion}$ is the potential at the ion's position in
the lattice.  In a hypothetical situation when a neutral molecule is
replaced with a foreign molecule which, subject to the potential
$\overline{\phi}_i$, has charge distribution $\overline{\rho}_i$, the
source term is zero, and such a molecule will not disturb the
translationally-invariant self-consistent solution.

With ions present, the problem has no translational symmetry, and the
number of self-consistent equations is infinite.  We consider an
imaginary cluster within an infinite lattice that includes all the
unit cells within a certain distance $R$ from the origin.  Some dozens
of molecules are required for the cluster to resemble a sphere.  We
set $\delta\rho_i^a=\delta\bbox{\mu}_i^a=0$ for the molecules outside
the cluster, and solve the self-consistent Eqs.~(\ref{SC3}) and
(\ref{SC4}) for $\delta\rho_i^a$, $\delta\bbox{\mu}_i^a$,
$\delta\phi_i^a$, and $\delta\bbox{F}_i^a$ within.  This corresponds
to an infinite lattice in which only the charges within the cluster
are allowed to relax.  Molecules outside retain the charge
distribution of the neutral lattice.

Setting $\delta\rho_i^a=\delta\bbox{\mu}_i^a=0$ does not make
$\delta\phi_i^a$ and $\delta\bbox{F}_i^a$ zero outside the cluster.
Nevertheless, we can write an expression for the polarization energy
of ions that does not contain self-consistent potentials and fields
outside the cluster.  Subtracting Eq.~(\ref{Etot1}) for the lattice
with ions from the similar expression for the
translationally-invariant lattice we obtain

\begin{eqnarray}
\Delta E_{\rm tot}&=&\frac{1}{2}\sum_a\sum_i(\delta\rho_i^a\phi_i^{a(0)}-
   \delta\bbox{\mu}_i^a\bbox{F}_i^{a(0)})\nonumber\\
  &+&\frac{1}{2}\sum_{ion}\sum_i\Delta\rho_i^{ion(0)}\phi_i^{ion}.
\label{DE}
\end{eqnarray}
 The first sum runs over all molecules $a$ including the ions.  The
second sum over the ions appears because both $\rho$- and
$\phi-$components of the bilinear expression Eq.~(\ref{Etot1}) are
different at $a=ion$.  The potentials $\phi_i^{a(0)}$ and fields
$\bbox{F}_i^{a(0)}$ in the unrelaxed translationally-invariant lattice
are evaluated using Eqs.~(\ref{SC2t}) with $\rho_i^b=\rho_i^{(0)}$ and
$\bbox{\mu}_i^b=0$.  Thus, Eq.~(\ref{DE}) gives the energy of a set of
ions in an {\em infinite} lattice in which molecules beyond the
imaginary cluster are not allowed to relax.  The polarization energy
of the set of ions in an infinite lattice is obtained as
$R\rightarrow\infty$.

\subsection{$P_\pm$ in anthracene and PTCDA}

We start with a single ion, when Eq.~(\ref{DE}) yields either $P_+$ or
$P_-$ in Eq.~(\ref{Et}).  The cluster of radius $R$ is centered on the
unit cell that contains the anion or cation.  Since clusters are
defined in terms of unit cells, we know the number of molecules $M(R)$
and $Mv_c/N_c=4\pi R^3/3$ relates $R$ to the molecular volume
$v_c/N_c$ in the crystal.  The polarization energy $P_++P_-$ for ions
at infinite separation is shown in Fig.~3 as a function of $M^{-1/3}$
for anthracene and PTCDA crystals.  The ``charges only'' points refer
to $\widetilde\alpha=0$ and polarization due entirely to charge
redistribution; the other points are based on the B3LYP values of
$\alpha$ (Table~I).  The largest clusters shown in Fig.~3 contain
$M=2000$ molecules, which corresponds to a cluster diameter of
$2R=114$ \AA\ for PTCDA and 97 \AA\ for anthracene.

\centerline{\epsfig{file=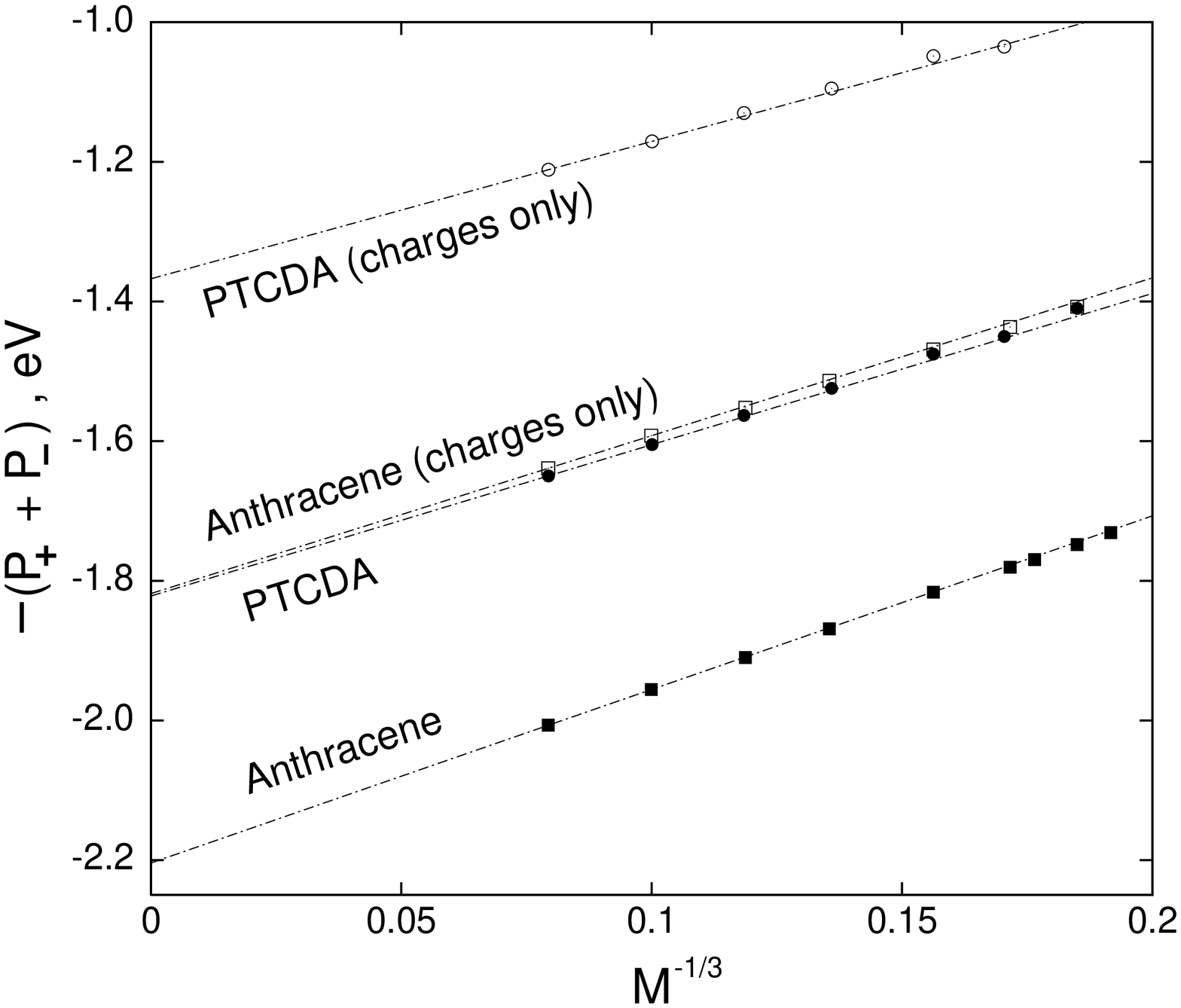,width=7.3cm,angle=-90}}
 \vskip 0.1 in
 {\small {\bf Fig.~3} Convergence of $P_++P_-$ for anthracene and
PTCDA with $M^{-1/3}$, which is proportional to the inverse radius
$R^{-1}$ of the cluster.  Straight lines are linear fits.  Open
symbols show the ``charges only'' results with $\widetilde\alpha=0$.}
 \vskip 0.1 in

The polarization energy decreases with cluster size as more degrees of
freedom for charge relaxation are added.  At large $R$ the missing
part due to the molecules outside the cluster can be thought of as the
polarization energy of a charge in the center of a cavity of radius
$R$ in a continous dielectric medium.  Such energy is linear in $1/R$.
Linear extrapolation in Fig.~3 gives $P_++P_-=-2.204$ eV for
anthracene and $-1.822$ eV for PTCDA crystals.  The smaller ion has
the greater stabilization.

The polarization energy of a charge in a spherical cavity in an
anisotropic medium has been evaluated by Bounds and Munn\cite{bounds}

\begin{equation}
P_\pm(\infty)-P_\pm(R)=
 -\frac{e^2}{2R}\left(1-\frac{1}{\kappa_{\text{eff}}}\right).
\label{DP}
\end{equation}
 The effective dielectric constants $\kappa_{\text{eff}}$ is expressed
in terms of the principal values $\kappa_1<\kappa_2<\kappa_3$ of the
dielectric tensor

\begin{equation}
\kappa_{\text{eff}}=\frac{\sqrt{\kappa_2(\kappa_3-\kappa_1)}}
 {F\left(\arctan\sqrt{
   \frac{\kappa_3-\kappa_1}{\kappa_1}},
   \sqrt{\frac{\kappa_3(\kappa_2-\kappa_1)}
              {\kappa_2(\kappa_3-\kappa_1)}}
 \right)}
\label{epseff}
\end{equation}
 where $F(\phi,k)$ is the elliptic integral of the first kind.
Equations (\ref{DP}) and (\ref{epseff}) determine the slope of the
asymtotic behavior of $P_\pm$ in Fig.~3.  As a consistency check we
computed the slope using the dielectric tensor obtained in Sec.~VIII.
The values 2.482 eV\AA\ for anthracene and 2.167 eV\AA\ for PTCDA are
within 3\% of the slope of the straight line in Fig.~3 drawn through
the last two calculated points.  The ``charges only'' slopes are also
within 3\% of the dielectric constants based on $\widetilde\alpha=0$.

Submolecules necessarily yield $P_+$ = $P_-$, since ions are assumed
to have equal and opposite charges and the neutral lattice contains
neither charges nor dipoles.  The anthracene result\cite{bounds} is
$P_\pm=1.42$ eV for three points at the centers of rings and an
effective $\alpha$ based on the static dielectric tensor of the
crystal, while experimental molecular $\alpha$ yields 1.51 eV for
three submolecules.  The electronic polarization, $P_++P_-\sim2.8$ eV,
substantially exceeds our 2.20 eV.  For PTCDA, 11 submolecules and
$\alpha$ similar to Table~I lead to\cite{petelenz} $P_++P_-=2.14$ eV,
which is again greater than our 1.82 eV.  While the results clearly
depend on the inputted $\alpha$, the polarization energy of single
ions found via charge redistribution is reduced compared to
submolecules, especially when only one or a few are used.  The
difference is even greater when the ion's charges $\rho_i^{ion(0)}$
are frozen at the gas-phase values.  This decreases $P_++P_-$ by
10---20 \% in these systems.  Although anthracene and PTCDA ions are
at inversion centers, their atoms are not and thus experience local
fields that are related by inversion.  By contrast, submolecule
charges are fixed at the outset and the field vanishes by symmetry at
the center of the molecule.

The transport gap, Eq.~(\ref{Et}), of molecular crystals is directly
related\cite{kahn} to photoelectron (PES) and inverse photoelectron
spectra (IPES) on surfaces, which yield adiabatic $P_++P_-$ that
include\cite{silinsh} intramolecular relaxation, but not lattice
relaxation.  The inferred\cite{kahn} $P_++P_-\sim1.7$ eV for PTCDA
films is quite consistent with the calculated 1.82 eV in the crystal.
The importance of $E_t$ for electronic organic devices and recent thin
film data were the motivation for the accurate calculation of
electronic polarization in the well-defined limit of zero overlap.
The systems of interest\cite{forrest,dodabalapur,batlogg,kahn} have
mobilities of 0.1---1.0 cm$^2/$Vs at room temperature, which is high
for organics and indicates that overlap corrections will have to be
included.

Partial charges and induced dipoles in the neutral lattice lead to
$P_+\neq P_-$.  The individual components are shown in
Fig.~4 and go as $M^{-1/3}\propto1/R$.  The anion and cation slopes
are equal at large $R$, in accord with Eq.~(\ref{DP}).  $P_+$ and
$P_-$ are almost identical for anthracene and strikingly different for
PTCDA.  In the smallest cluster, which contains only the anion or
cation, the ion interacts with the charges $\overline\rho_i$ and
dipoles $\overline{\bbox{\mu}}_i$ of the neutral lattice.

Finite $P_+(M=1)$ and $P_-(M=1)$ are the energies of the cation and
anion in the unrelaxed neutral lattice.  They are nonzero due to
fields in the neutral lattice.  Without relaxation of the ion itself,
$P_\pm(M=1)$ is given by the second term of Eq.~(\ref{DE}) with
potentials $\phi_i^{ion}=\overline\phi_i+\phi_i^{(0)}$.  The relaxed
ion in the field of the neutral lattice has charge distribution
$\rho_i^{ion*}$, and $P_\pm(M=1)$ is given by Eq.~(\ref{DE}) with
$\delta\rho_i^a=\rho_i^{ion*}$ for $a=ion$ and 0 otherwise.  We have
assumed $\widetilde\alpha^{ion}=\widetilde\alpha$ for simplicity; more
generally, nonzero
$\Delta\widetilde\alpha=\widetilde\alpha^{ion}-\widetilde\alpha$ will
introduce a source term
$\bbox{\mu}_i^{ion*}=\Delta\widetilde\alpha_i\overline{\bbox{F}}_i$ in
Eq.~(\ref{SC3b}).

The large PTCDA contributions at $M=1$ do not cancel exactly because
the anion and cation charges are not precisely equal and opposite.
Approximate electron-hole symmetry for the $\pi$-system of anthracene
ensures almost equal and opposite charges.  Our treatment gives two
contributions to $P_\pm$, an initial interaction at $M=1$ that does
not arise for submolecules and a relaxation or polarization of the
lattice that remains almost the same for the anion and cation.  Such a
distinction may be useful in future work.

The charge distribution in PTCDA is such that positive atomic charges
are closer to the molecular centers.  The resulting quadrupole moments
of molecules create an average positive potential at each molecule in
the neutral lattice.  The $\pi$-electron density above and below the
molecular plane also generates a quadrupole as discussed by Silinsh
and Capek.\cite{silinsh}

\centerline{\epsfig{file=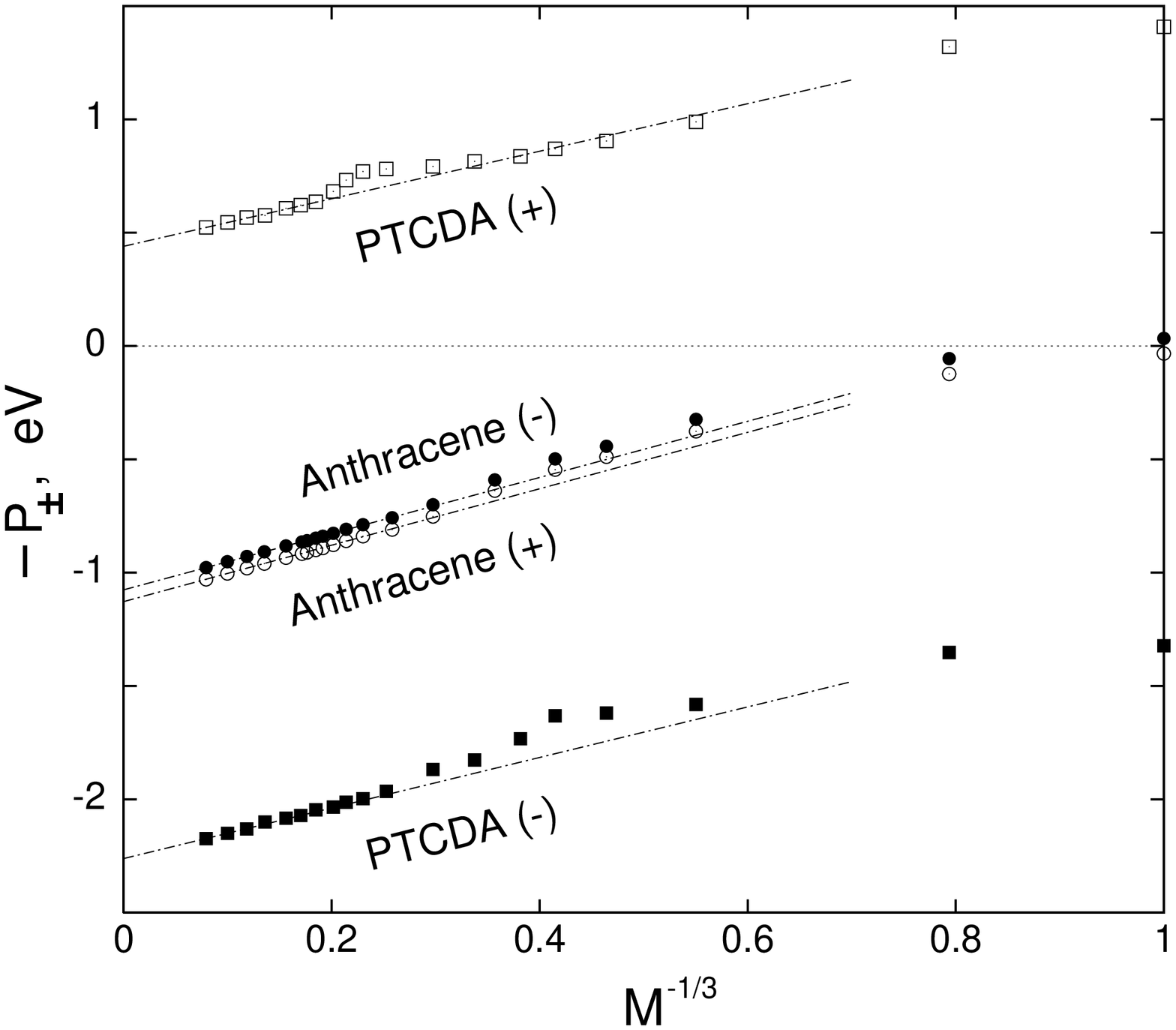,width=7.8cm,angle=-90}}
 \vskip 0.1 in
 {\small {\bf Fig.~4} Convergence of $P_+$ and $P_-$ for anthracene
and PTCDA with with $M^{-1/3}$.  Straight lines are linear fits.  The
values at $M=1$ are discussed in the text.}
 \vskip 0.1 in

In fact, the quadrupole contribution to $P_+$ and $P_-$ depends on the
macroscopic shape of the sample.  It is proportional to $\int d^3r
(1/r^3)$, which gives finite contribution from the remote parts of the
sample.  The contribution to the potential is constant on the scale of
the unit cell, because the corresponding contribution to the field,
$\propto\int d^3r (1/r^4)$, vanishes.  Thus, the individual quantities
$P_+$ and $P_-$ are not well defined, but the shape-dependent
contribution cancels exactly in the sum, $P_++P_-$, which enters
Eq.~(\ref{Et}) for the gap.

The quantities $I-P_+$ and $A+P_-$ can be viewed as the ionization
potential and the electron affinity of the solid.  We see that they
depend on the macroscopic shape of the sample due to quadrupolar
corrections.  In general, the polarization energy of an abitrary set
of charges in a crystal lattice depends on the shape of the
macroscopic sample, unless the total charge is zero.

The interpretation of $P_+$ and $P_-$ is, however, of considerable
interest because PES and IPES spectra are related\cite{kahn} to
$I-P_+$ and $A+P_-$, respectively.  Direct comparison, therefore,
requires a surface calculation.  Compilations\cite{pope,silinsh} of
$I-P_+$ for organics, while admittedly approximate, clearly point to
$P_+>P_-$ and to the physical meaning of individual polarization
energies.

\subsection{Ion pairs in anthracene and PTCDA}

Polarization effects modify the interaction between charge carriers.
We compute $V({\bf r})$ in Eq.~(\ref{Epair}) by replacing two neutral
molecules in the lattice with a cation and anion.  The cation is at
the origin and the anion's center has crystallographic coordinates r =
(a,b,c) given in Table~II.  As in the previous section, we consider an
imaginary cluster of radius $R$ that contains both ions, solve the
self-consistent Eqs.~(\ref{SC3}) and (\ref{SC4}), and evaluate the
energy $E_{\rm pair}$ using Eq.~(\ref{DE}).  We repeat with larger $R$
until $V({\bf r})$ converges.

\centerline{\epsfig{file=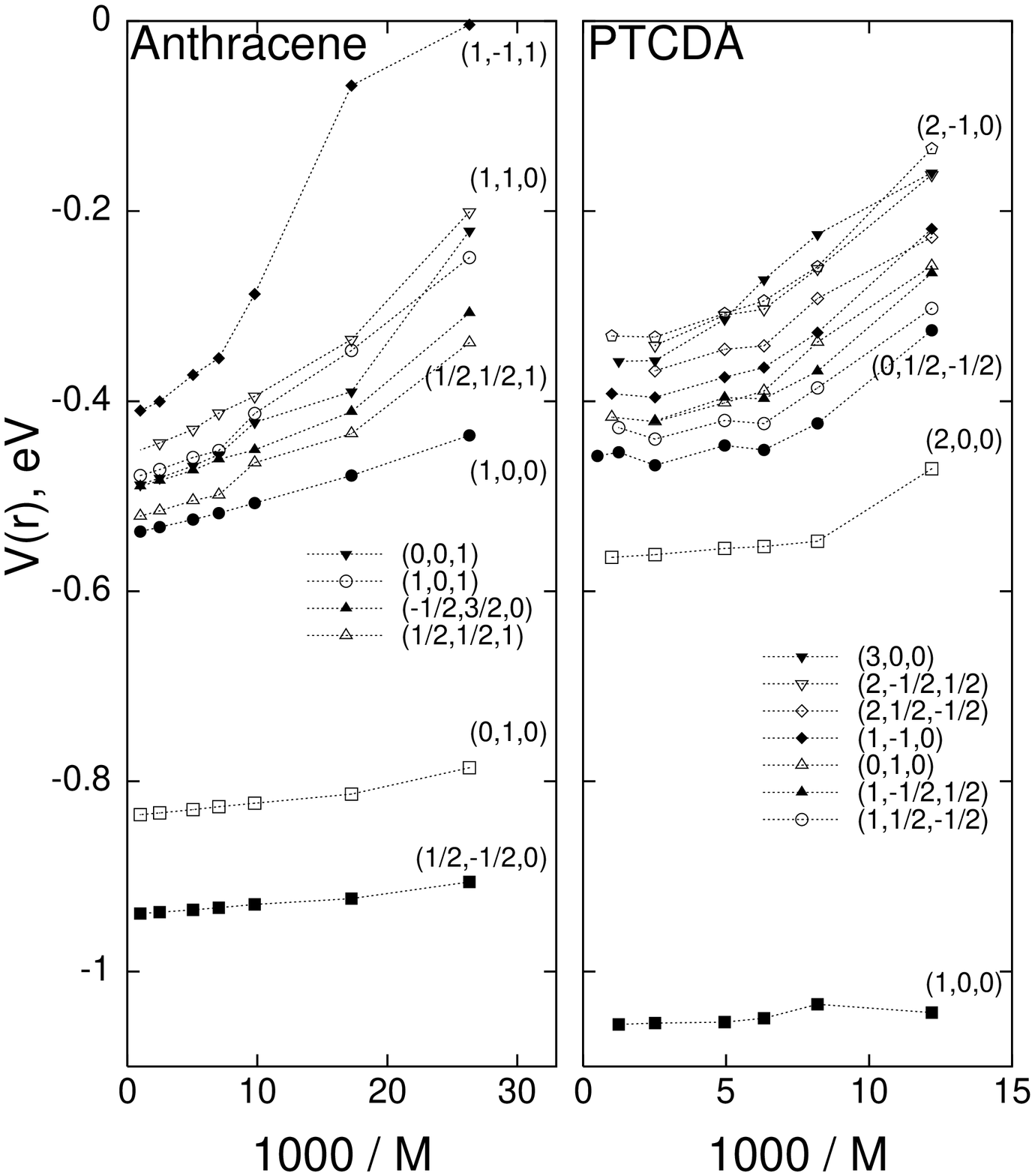,width=9cm}}
 {\small {\bf Fig.~5} Interaction energy $V({\bf r})$,
Eq.~(\ref{Epair}), for various ion pairs in clusters of $M$
molecules.}
 \vskip 0.1 in

Figure 5 shows $V({\bf r})$ as a function of $1/M$, which is
proportional to the inverse cluster {\em volume}, for various ion
pairs.  Since the pair is neutral, the $1/R$ contribution given by
Eq.~(\ref{DP}) vanishes, and the asymptotic behavior is linear in
$1/R^3$.  It represents the polarization energy of a dipole in the
center of spherical cavity in a dielectric medium.

The extrapolated values of $V({\bf r})$ for various pairs are
presented in Table~II, which also lists the distances between centers
and identifies pairs using the crystallographic notation of
Ref.~\onlinecite{karl}.  The lowest charge-transfer (CT) exciton in
PTCDA (Fig.~5) corresponds to neighbors in the stack.  The next CT
state is the second neighbor along the stack, which is closely
followed in energy by other configurations for neighboring molecules
in different stacks, as shown in Fig.~2.  In anthracene the lowest CT
state corresponds to the closest neighbor.

Bounds and Siebrand\cite{siebrand} obtained $V({\bf r})$ for
anthracene using a single point per molecule and experimental
polarization data.  Their 0.78 and 0.58 eV binding for the lowest CT
states are less than our 0.922 and 0.821 eV; the large difference for
the second neighbors is due to in-plane polarization and the
inadequacy of a single point charge for long molecules that are
end-to-end.  Anisotropic $\alpha$ and charge redistribution in
anthracene produce several instances (e.g. at 9.894 and 11.172 \AA\ )
where $V({\bf r})$ is not monotonic in $r$.

 \vskip 0.2 in
 {\small {\bf Table~II} Energies $V({\bf r})$, in eV, of
charge-transfer states within zero-overlap approximation for several
separations ${\bf r}$.  The cation and anion are at $(0,0,0)$ and
$(a,b,c)$ respectively.}

\noindent
\begin{center}
\begin{tabular}{lrrr}
\\
\tableline
\tableline
 Pair $(a,b,c)$\ \ \  &
 \ \ \ \ \ $r$, \AA\ \ \ \ \ \ &
 \ \ \ \ $V({\bf r})$\ \ \ \ \  &
 \ \ \ \ \ $V_{\rm appr}({\bf r})^a$ \\
\tableline
\\
\multicolumn{3}{l}{\bf \ \ \ \ Anthracene} & \\
 $(\frac{1}{2},-\frac{1}{2},0)$  &  5.228 & -0.92 & -0.79 \\
 $(0,1,0)$                       &  6.016 & -0.82 & -0.76 \\
 $(1,0,0)$                       &  8.553 & -0.55 & -0.50 \\
 $(1,0,1)$                       &  9.458 & -0.51 & -0.50 \\
 $(\frac{1}{2},\frac{1}{2},1)$   &  9.894 & -0.54 & -0.57 \\
 $(-\frac{1}{2},\frac{3}{2},0)$  &  9.986 & -0.51 & -0.47 \\
 $(1,1,0)$                       & 10.456 & -0.49 & -0.42 \\
 $(0,0,1)$                       & 11.172 & -0.51 & -0.56 \\
 $(1,-1,1)$                      & 11.209 & -0.50 & -0.42 \\
\\
\multicolumn{3}{l}{\bf \ \ \ \ PTCDA} & \\
 $(1,0,0)$                       &  3.726 & -1.06 & -0.75 \\
 $(2,0,0)$                       &  7.453 & -0.57 & -0.46 \\
 $(0,\frac{1}{2},-\frac{1}{2})$  & 10.558 & -0.46 & -0.48 \\
 $(1,-\frac{1}{2},\frac{1}{2})$  & 10.751 & -0.43 & -0.40 \\
 $(3,0,0)$                       & 11.179 & -0.38 & -0.32 \\
 $(1,\frac{1}{2},-\frac{1}{2})$  & 11.624 & -0.43 & -0.46 \\
 $(0,1,0)$                       & 11.998 & -0.42 & -0.46 \\
 $(2,-\frac{1}{2},\frac{1}{2})$  & 12.144 & -0.36 & -0.32 \\
 $(1,-1,0)$                      & 12.563 & -0.39 & -0.42 \\
 $(2,\frac{1}{2},-\frac{1}{2})$  & 13.658 & -0.38 & -0.37 \\
 $(2,-1,0)$                      & 14.124 & -0.33 & -0.32 \\
\tableline
\tableline
\multicolumn{4}{l}{$^a$ Eq.~(\protect\ref{Vapprox})}
\\
\end{tabular}
\end{center}

The point-charge approximation\cite{shen_forrest} gives almost 2.0 eV
of binding for PTCDA neighbors in a stack, twice the 1.05 eV in
Table~II, while the binding\cite{petelenz} is 0.99 eV for 11
submolecules.  Point charges are poor approximations for large
molecules with interplanar separation of 3.4 \AA.  Charge
redistribution and partial charges provide a direct way to compute
such electrostatic interactions.

We may consider $V({\bf r})$ for a cation-anion pair in a continous
anisotropic medium with dielectric tensor $\kappa$.  We describe each
ion with fixed partial charges $\rho_i$ and write an electrostatic
expression in terms of the double sum over the atoms of the cation and
anion.  Since $\Delta E_{\rm tot}$ is defined relative to the neutral
lattice, charge differences
$\Delta\rho_i^{ion}=\rho_i^{ion}-\overline\rho_i$ appear in $V({\bf
r})$:

\begin{equation}
V_{\rm appr}({\bf r})=
 \sum_{ij}\frac{\Delta\rho_i^+\Delta\rho_j^-}
  {[det(\kappa)\kappa^{-1}_{\alpha\beta}r_{ij}^\alpha r_{ij}^\beta]^{1/2}}.
\label{Vapprox}
\end{equation}
 Here $r_{ij}^\alpha$ are the components of ${\bf r}_{ij}={\bf
r}_i^+-{\bf r}_j^-$, and $det(\kappa)=\kappa_1\kappa_2\kappa_3$ is the
determinant of the dielectric tensor.\cite{landau} At large ${\bf r}$,
Eq.~(\ref{Vapprox}) reduces to point charges, since
$\sum_i\overline\rho_i=0$ ensures vanishing interactions in the
neutral lattice.  The lowest-order corrections to Eq.~(\ref{Vapprox})
are induced dipoles due to the other charge.

Table~II lists $V_{\rm appr}({\bf r})$ values based on
Eq.~(\ref{Vapprox}) and the dielectric tensors obtained in the next
Section.  Gas-phase charges and dielectric data are a simple and
reasonably accurate approximation to the self-consistent calculation.
 
We find comparable $V({\bf r})\sim-0.5$ eV for second neighbors in
PTCDA stacks and first neighbors in different stacks (Fig.~2).  Mazur
and Petelenz\cite{petelenz} report instead that, by a small margin, an
interstack neighbor has binding 1.02 eV that even exceeds the first
neighbor.  They emphasize the competition between large in-plane
polarization for neighbors in adjacent stacks and Coulomb interactions
in the same stack.  Such trade-offs are seen in Table~II for both
anthracene and PTCDA, although not as strongly as in
Ref.~\onlinecite{petelenz}.  They compute $E_{\rm pair}$ for
fractional charges at 11 submolecules and then find the Coulomb
interaction between the anion and cation using L\"owdin charges in a
6-31G basis.  While it is inconsistent to use different charges for
polarization and direct interactions, it is natural to prefer atomic
charges for the anion-cation interaction to arbitrarily placed partial
charges.

We also find (1,0,0) to be the lowest CT state by $\sim0.3$ eV on
using, as in Ref.~\onlinecite{petelenz}, 11 fractional charges and
partial polarizabilities for both the polarization and cation-anion
interaction.  Hence (1,0,0) is the lowest CT state in PTCDA and,
although important, the greater in-plane polarizability does not
stabilize neighbors in different chains below (2,0,0).

\section{Dielectric tensor}

We now summarize the calculation of the dielectric tensor by
considering a sample in a uniform electric field.  These results have
been reported previously.\cite{ours} The procedure is similar to the
polarization energy of the neutral lattice in Sec.~V.

Formally, an applied field breaks translational invariance, since the
electrostatic potential is unbound.  Nevertheless, we can add an
appropriate constant to the potential in each unit cell and restore
translational symmetry, without effect on charge distribution.  This
follows for zero overlap because space can be subdivided such that
each molecule feels its own potential $\phi({\bf r})$.  The charge
distribution does not change when a constant is added to all $\phi_i$,
according to Eq.~(\ref{Picond}).

We add $E_0^\alpha$ to the right-hand side of Eq.~(\ref{SC2tb}), and
the corresponding term $-r_i^{a\alpha}E_0^\alpha$ to the right-hand
side of Eq.~(\ref{SC2ta}).  We then solve $4NN_c$ self-consistent
equations (\ref{SC2t}) and (\ref{SC1}), and obtain the total dipole
moment $\bbox{\mu}^a$ of each molecule $a$ in the unit cell using
Eq.~(\ref{mu}).  The total dipole moment of the unit cell
$\bbox{\mu}=\sum_a\bbox{\mu}^a$ determines the polarization
$\bbox{P}=\bbox{\mu}/v_c$, where $v_c$ is the unit-cell volume.
Repeating this procedure with $\bbox{E}_0$ directed along each of the
coordinate axes we find the tensor $\zeta$ that relates $\bbox{P}$ to
$\bbox{E}_0$:

\begin{equation}
\bbox{P}=\zeta\bbox{E}_0.
\label{zeta}
\end{equation}
 Alternatively, one may choose to differentiate the self-consistent
equations with respect to $E_0^\alpha$ to find an explicit expression
for $\zeta$.

The susceptibility tensor $\chi$ is defined by the relation
$\bbox{P}=\chi\bbox{F}$, where $\bbox{F}$ is the total average
macroscopic field that is created by external sources as well as the
polarized solid itself.\cite{landau} According to the Lorentz
relation, a dielectric sphere with uniform polarization $\bbox{P}$
creates a field $-4\pi\bbox{P}/3$ in its center.  The spherical shape
is consistent with the definition of ${\cal V}({\bf r})$ in
Eq.~(\ref{Vcal}).  Thus,

\begin{equation}
\bbox{F}=\bbox{E}_0-\frac{4\pi}{3}\bbox{P}.
\label{lorentz}
\end{equation}
 Using Eq.~(\ref{lorentz}) we eliminate $\bbox{E}_0$ in favor of
$\bbox{F}$ and using $\kappa=1+4\pi\chi$ we obtain the dielectric
tensor

\begin{equation}
\kappa=\frac{1+(8\pi/3)\zeta}{1-(4\pi/3)\zeta}
\label{kappa}
\end{equation}

 \vskip 0.2 in
 {\small {\bf Table~III} Experimental (averaged) and calculated
components of dielectric tensor of anthracene and PTCDA}

\noindent
\begin{tabular}{lcccc}
\\
\tableline
\tableline
Inputs & $\kappa_1$ & $\kappa_b$ & $\kappa_3$ & $\theta^o$ \\
\tableline
\\
\multicolumn{2}{l}{\bf \ \ \ \ Anthracene} & & \\
Experiment\protect\cite{forrest_apl,karl_eps,munn_eps,winchell}
                         & 2.49(10)  & 3.07(10)  & 4.04(20) & 28(2) \\
B3LYP/6-311++G**         & 2.23      & 2.91      & 4.03     & 31.6 \\
Charges only ($\widetilde\alpha=0$)
                         & 1.36      & 2.39      & 3.90     & 34.5 \\
\\
\multicolumn{2}{l}{\bf \ \ \ \ PTCDA} & & \\
Experiment\cite{forrest_apl}
                         & 1.9(1)    & 4.3(2)    & 4.6(2)   &      \\
                         & 1.85      & 4.07      & 4.07     &      \\
B3LYP/6-311++G**         & 1.96      & 3.98      & 4.02     &      \\
Charges only ($\widetilde\alpha=0$)
                         & 1.01      & 3.74      & 3.81     &      \\
\tableline
\tableline
\\
\end{tabular}

Table III lists the principal components of $\kappa$ in anthracene
crystals and PTCDA films.  $\kappa_2$ is along the b axis; $\theta$ is
the angle in the ac plane between $\kappa_1$ and $a$.  The anthracene
data\cite{forrest_apl,karl_eps,munn_eps,winchell} are averages over
independent measurements of the dielectric tensor and refractive
indices.  The calculated values are for $\alpha$ in Table~I, and
additional $\alpha$ inputs for larger bases are reported in
Ref.~\onlinecite{ours}.  The results agree with the available
experimental data, which has an accuracy of a few percent.  While
$\widetilde\alpha=0$ accounts for $\kappa_3$, the largest component,
the quantitative importance of accurate $\alpha$ is clearly seen in
Table~III.

Equation (\ref{kappa}) for $\kappa$ is strictly based on neutral
molecules in a translationally-invariant crystal.  The energies of
molecular ions were found instead within clusters of radius $R$.  As
noted above, Eq.~(\ref{DP}) reproduces the $M^{-1/3}$ slope in Figs.~3
and 4 within 3\%.  This demonstrates the internal consistency of the
procedure.

\section{Discussion}

In the limit of zero overlap, molecules in organic crystals are
quantum systems with purely classical electrostatic interactions.  We
have developed a self-consistent approach that treats each molecule
quantum-mechanically, subject to the external fields of all other
molecules.  We have found that such external fields can be treated
perturbatively to a good accuracy, expanding the solution of
Schr\"odinger equation for each molecule near the gas-phase solution.
Self-consistent analysis of large systems, with over $10^5$ atoms, is
straightforward.

The self-consistent procedure captures an important effect: the
redistribution of charge in molecules subject to (nonuniform) external
fields.  Direct description of charge redistribution avoids the
ambiguity that accompanies partitioning the molecular polarizability
over a number of polarizable points.

As outlined in Section III, we discretize the molecular charge
distribution $\rho({\bf r})$ by assigning a partial charge $\rho_i$
and induced dipole $\bbox{\mu}_i$ to every atom $i$.  The atom-atom
polarizability tensor $\Pi_{ij}$ then governs charge redistribution in
external fields.  The tensor $\Pi_{ij}$ is conveniently found
quantum-mechanically at a semiempirical level, such as INDO/S, which
is well suited for introducing potentials or site energies.  The
correction $\widetilde\alpha=\alpha-\alpha^C$ is distributed over the
atoms, and as in previous theory,\cite{silinsh,munn} we still face the
familiar problems associated with submolecules.  The present approach,
however, allows for a certain compensation of error: the partitioning
now only involves the small correction $\widetilde\alpha$, which is
about 10---20\% of the actual molecular polarizability $\alpha$ in
large conjugated molecules.

The electrostatic potential $\phi({\bf r})$ created by a molecule with
charge distribution $\rho({\bf r})$ is given by Eq.~(\ref{phitot}).
While atomic charges cannot be defined uniquely, charges or dipoles
that reproduce $\phi({\bf r})$ accurately at intermolecular distances
in crystals would be completely satisfactory.  Thus, {\em ab-initio}
$\rho({\bf r})$ generate $\phi({\bf r})$ outside the molecule, which
provides a basis\cite{chirlian} for assigning discrete charges.  This
suggests an interesting possibility of introducing gas-phase atomic
dipoles $\bbox{\mu}_i^{(0)}$ along with the gas-phase atomic charges
$\rho_i^{(0)}$ in Eq.~(\ref{SC1b}), as long as they improve the
description of fields created by a neutral molecule.

Another possible extention is to introduce atomic quadrupoles to
represent $\pi$-electron density above and below the plane.  This will
result in an additional $10N$ scalar equations per molecule.  In
practice, only the $q_{NN}$ component normal to the conjugation plane
is needed for $\pi$-electrons.  Classical multipoles lead to
complications even in the limit of no overlap.  These are corrections,
however, to charges and induced dipoles whose fields $\bbox{F}_i$ we
have found, and perturbation theory may well be sufficient.

We have already contrasted our method to the existing approach in
which a molecule is represented by a number of polarizable points,
termed submolecules.  The qualitative differences are the
electrostatic energy of the neutral lattice and the energy
$P_\pm(M=1)$ of an ion in the unrelaxed lattice in Fig.~4.  Charge
redistribution avoids entirely the number and location of the
submolecules.  Since the actual $\alpha$ enters in either case, the
numerical results for $P_++P_-$ in Section~VI and for $V({\bf r})$ in
Table~II do not differ greatly, especially in comparison to
calculations with many submolecules.

Crystalline organic films that function as electron or hole conductors
are of particular interest for organic
electronics.\cite{forrest,dodabalapur,batlogg} Although overlap is
then finite, it should be considered as a correction to much greater
polarization energies.  We have a systematic method for computing the
electronic part of $P_++P_-$, while the total polarization energy in
Eq.~(\ref{Et}) is about $10$\% larger due to the lattice
contributions, even before overlap or charge transport is introduced.
Reliable comparisons of the large electronic part are the first step.
We comment on two issues.

First, the greater stabilization of separated ions has major
implications on the binding energy $|V({\bf r})|$ of neighboring ions.
Since the energy of the lowest CT in PTCDA\cite{hennessy} is
comparable to the lowest Frenkel exciton, $|V({\bf r})|$ is closely
related to discussions and debates about the magnitude of exciton
binding energies.\cite{saraciftci} Lattice (molecular) relaxation
about a separated anion and cation is probably in the range of 200 meV
each, while relaxation about an adjacent ion pair is less than 100
meV; the lattice contribution decreases $|V({\bf r})|$.  Finite
overlaps and transfer integrals in the 50---100 meV range are typical
for hopping transport and stabilize polarons in Holstein models.  CT
states, by contrast, have vanishing bandwidth or mobility because
two-electron transfers are needed.  Overlap also preferentially
stabilizes individual ions and hence reduces $|V({\bf r})|$.  The
lowest CT state of PTCDA or anthracene has electronic binding of
$\sim1$ eV in Table~II that could be reduced substantially due to
lattice relaxation or by overlap in systems with mobile charges.

The second issue addresses novel features of self-consistent atomic
charges.  As noted above, charge redistribution of isolated ions
increases $P_+$ or $P_-$ by about 10\% even when ions are at inversion
centers.  The lower symmetry of ion pairs leads to more extensive
charge redistribution of mutual polarization that is automatically
included in our treatment through $\Pi_{ij}$ for ions.  Charge
redistribution on ions is important.  Its contribution can be
estimated by self-consistent calculation with ionic charges and
dipoles set to gas-phase values.  Specific contributions are found by
comparing two self-consistent solutions.

In summary, we have implemented a general approach to electronic
polarization of organic molecular crystals in the limit of zero
intermolecular overlap.  Redistribution of partial atomic charges is
governed by the atom-atom polarizability tensor $\Pi_{ij}$, which
provides a quantum mechanical basis for the description of electronic
polarization.  Partial charges replace the microelectronics of
postulated submolecules in previous treatments and introduce such new
features as electronic stabilization of the neutral lattice and charge
relaxation on ions.  Self-consistent atomic charges and induced
dipoles relate the molecular polarizability of anthracene or PTCDA to
the dielectric tensor and the energies of fixed ions and ion pairs in
the crystal.  The cluster approach used for ions is suitable for
surfaces and other systems with reduced symmetry, provided that all
molecular positions are specified.  Zero overlap provides a starting
point for the treatment of electronic polarization in organic systems
with mobile localized charge carriers.

\acknowledgements

It is a pleasure to thank R.A. Pascal, Jr., for discussions and
assistance with structural data and molecular polarizabilities;
A. Kahn, S.F. Forrest, M. Hoffmann, and J.M. Sin for discussions of
transport gaps, PES spectra, CT excitons and atomic charges; and
P. Petelenz and R.W. Munn for correspondence about induced dipoles.
This work was partly supported by the National Science Foundation
through the MRSEC program under DMR-9400362.

\end{multicols}
\end{document}